\documentclass[12pt]{article}

\usepackage[utf8]{inputenc} 
\usepackage[T1]{fontenc}    
\usepackage{hyperref}       
\usepackage{url}            
\usepackage{booktabs}       
\usepackage{amsfonts}       
\usepackage{nicefrac}       
\usepackage{microtype}      
\usepackage[inline]{enumitem}
\usepackage{ftnxtra}
\usepackage{fullpage}

\usepackage{graphicx}
\usepackage{subcaption}

\usepackage{setspace}
\providecommand{\keywords}[1]
{
  \small	
  \textbf{\textit{Keywords---}} #1
}

\title{Holistic generational offsets: Fostering a primitive online abstraction for human vs. machine cognition}

\author{
  Shaun D'Souza\footnote{Former affiliation} \quad Trevor Mudge \\
  Computer Science and Engineering \\
  University of Michigan \\
}

\date{}

\begin{document}

\maketitle

\begin{abstract}
We propose a unified architecture for next generation cognitive, low cost, mobile internet. The end user platform is able to scale as per the application and network requirements. It takes computing out of the data center and into end user platform. Internet enables open standards, accessible computing and applications programmability on a commodity platform. The architecture is a super-set to present day infrastructure web computing. The Java virtual machine (JVM) derives from the stack architecture. Applications can be developed and deployed on a multitude of host platforms. O(1) $\leftrightarrow$ O(N). Computing and the internet today are more accessible and available to the larger community. Machine learning has made extensive advances with the availability of modern computing. It is used widely in NLP, Computer Vision, Deep learning and AI. A prototype device for mobile could contain N compute and N MB of memory.
\end{abstract}


\section{Introduction}

The web ecosystem is rapidly evolving with changing business and functional models. Cloud platforms are available in a SaaS, PaaS and IaaS model designed around commoditized Linux based servers. 10 billion users will be online and accessing the web and its various content. The mobile and internet are ubiquitous today. The industry has seen a convergence around IP based technology. Additionally, Linux based designs allow for a system wide profiling of application characteristics.

\begin{figure}[!h]
  \caption{Host architecture.}
  \centering
  \includegraphics[width=0.3\columnwidth]{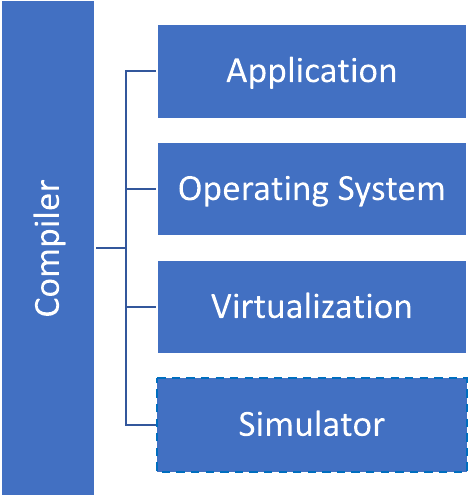}
  \label{fig:host}
\end{figure}

A virtualized architecture consists of Figure~\ref{fig:host}. The compiler is the glue logic for all the layers interfacing software to the underlying platform. Application performance was a primary determinant of system performance. Processor and Memory technology determine system performance~\cite{Kgil:2006:PUS:1168857.1168873}. With the advent of the Internet, computing performance is increasingly being utilized in the network. Applications are internet based and network connectivity is central to the platform. Network performance is a primary determinant of system. Existing internet connectivity are limited by technology capabilities Wi-Fi, 4G (Mbps).

\begin{figure}[!h]
  \caption{Conventional computer platform components. Compute coupled to memory and network.}
  \centering
  \includegraphics[width=0.4\columnwidth]{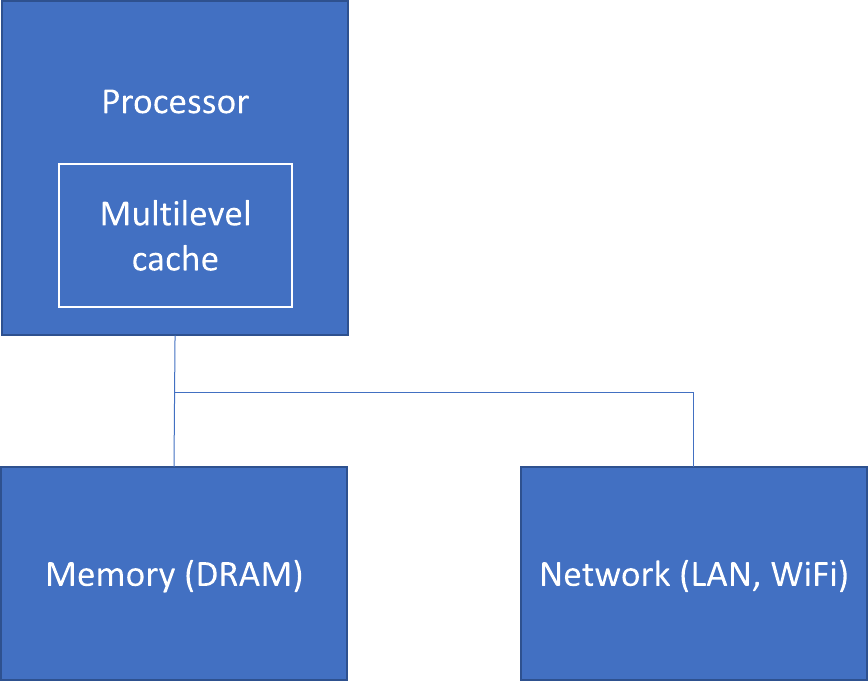}
  \label{fig:proc}
\end{figure}

Conventional Computer platform components
\begin{itemize}
\item Processor, Cache
\item Memory (DRAM)
\item Network (LAN, Wi-Fi)
\end{itemize}

Increasingly applications are internet based. Network performance is a primary determinant of system performance. Conventional computer was an in the box solution for desktop applications. The architecture was developed for high performance desktop applications. Processor, cache (GHz) coupled to high capacity Memory (DRAM). Figure~\ref{fig:proc} shows a conventional platform with Processor and cache coupled to Memory and Network.

Processor technology speeds are increasing faster than DRAM memory technology. Processors are designed to operate at a high frequency \textgreater 2 Ghz. Caches are coupled to the processor to facilitate execution at high speed. DRAM memory technology is designed for high density \textgreater 2 GB. As a result, the platform is not able to scale to meet the network performance and system performance requirements. 

Network applications rely of moving data from network to memory and the processor - Figure~\ref{fig:proc}. As a result, system performance is determined by bandwidth throughput in the memory. Internet applications are consuming increasing bandwidth and will use \textgreater 1 – 10 Gbps bandwidth in a mobile platform.

Processor and memory (DRAM) technology have evolved independently to increase system performance. Processors were designed to run at high speeds (\textgreater 2 GHz). Memory (DRAM) was designed for large capacity (\textgreater 2 GB)

Until recently it was not feasible to integrate multi-Mb memory in a processor. Caches were used in the processor with application residing in main memory DRAM. However, with technological advances it is now possible to integrate multi-Mb memory (SRAM) in a processor. This enables us to re-evaluate the system hierarchy with processor, memory and network - Figure~\ref{fig:xbar}.

Integrating large memory in the processor allows us to eliminate caches in the design. Desktop applications were constrained by technology performance~\cite{Lindholm:2014:JVM:2636992}. However recently we are seeing a saturation in application performance requirements. The Internet is today the platform for application enabling and the internet operational enablement is a driver of technology growth~\cite{berners1989information}.

\begin{figure}[!h]
  \caption{As-a-service cloud.}
  \centering
  \includegraphics[width=0.7\columnwidth]{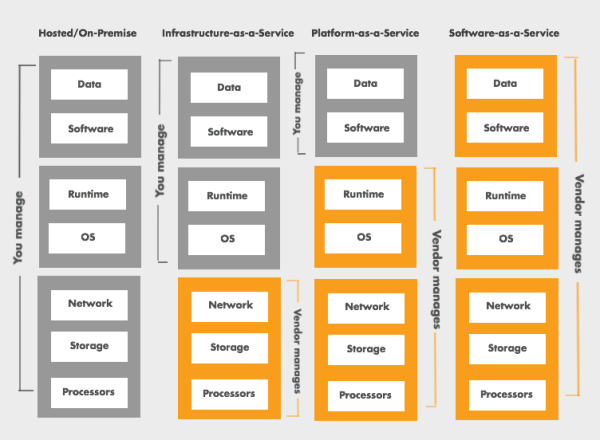}
  \label{fig:aas}
\end{figure}

{\bf Web N.} Open source has enabled the development of more efficient internet systems. Cloud computing combined with service-oriented architecture (SOA) and parallel computing have influenced application software development. Code is implemented in a variety of frameworks to ensure performance including OpenMP, MPI, MapReduce and Hadoop. Parallel programs present a variety of challenges including race conditions, synchronization and load balancing overhead. 

{\bf Virtualization.} This hosted infrastructure is complemented by a set of virtualization utilities that enable rapid provisioning and deployment of the web infrastructure in a distributed environment. Virtualization abstracts the underlying platform from the OS enabling a flexible infrastructure.

{\bf Open Source.} Additionally, the cloud ecosystem is supported by the Open Source community enabling an accelerated scale of development and collaboration~\cite{mahmood2013software}. Simulators are used to benchmark application systems.

\section{Differentiator}

We propose an architecture for the next generation enterprise including an end to end solution for the web infrastructure. This highlights the challenges in bringing billions of users online on a commodity platform. There is a large opportunity in enabling technology consumption for more than a billion users. 

\begin{itemize}
\item Web N, 10 Billion users, Intelligent machines, Turing test
\item Social media, enterprise mobility, data analytics and cloud
\item AI, Technology and enterprise, Virtualization, Open Source
\item Machine learning, compilers, algorithms, systems
\end{itemize}

OOP and Java have enabled enterprise system architecture. Java is an algorithms, web and enterprise centric programming language. It allows for deployment of applications on a host of platforms running a virtual machine. Write once, run anywhere (WORA). 3 billion mobile devices run Java. Enterprise applications provide the business logic for an enterprise. Architectures have evolved from monolithic systems, to distributed tiered systems, to Internet connected cloud systems today.

\begin{figure}[!h]
  \caption{Virtual memory.}
  \centering
  \includegraphics[width=0.25\columnwidth]{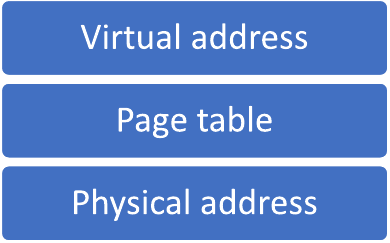}
  \label{fig:virtual}
\end{figure}

Figure~\ref{fig:virtual} shows the Virtual memory system on a host architecture and address translation to a physical address using the Page table in the Operating system.

\begin{figure}[!h]
  \caption{Conventional computer platform stack.}
  \centering
  \includegraphics[width=0.7\columnwidth]{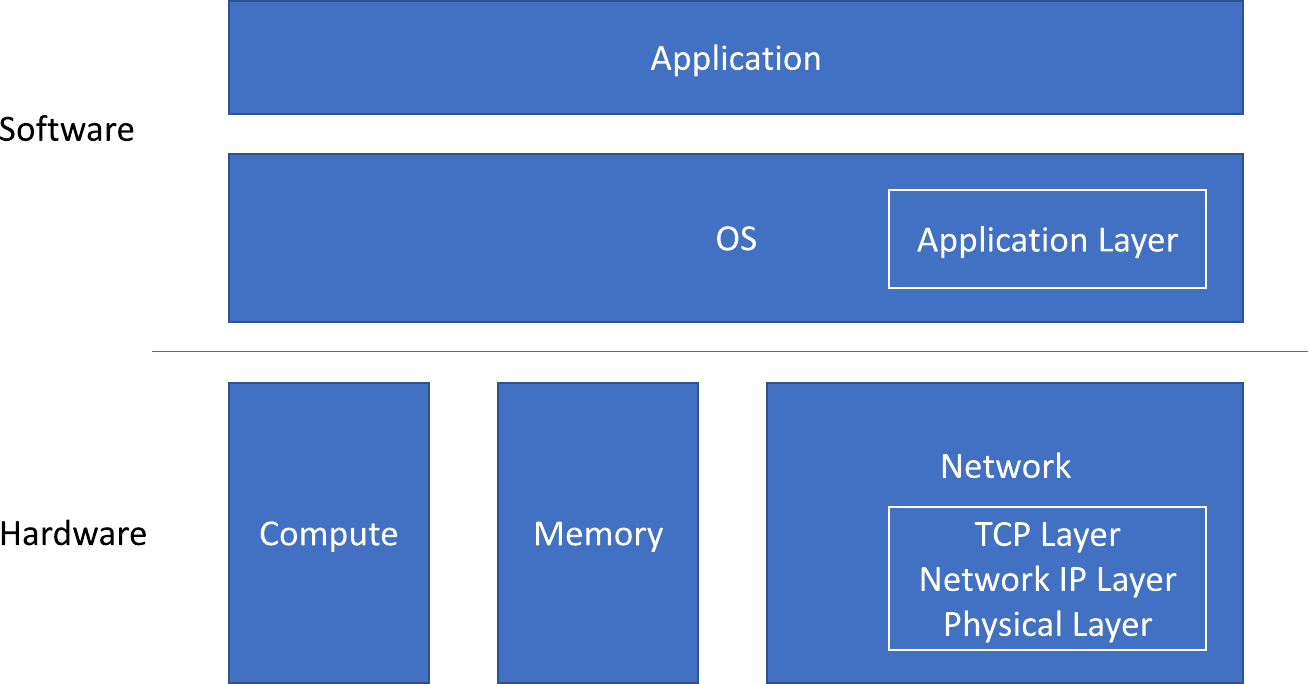}
  \label{fig:os_conv}
\end{figure}

Technology assumptions -
\begin{itemize}
\item Architect and design cloud applications to support multi-tenancy, concurrency management, parallel processing and service-oriented architecture supporting rest services.
\item Law of accelerating returns~\cite{kurzweil2004law} and Moore’s Law
\item Prices and margins, competition, converging global supply and demand, evolving business models
\end{itemize}

A conventional computer platform consists of - Figure~\ref{fig:os_conv}

Hardware:
\begin{itemize}
\item Processor and cache
\item Memory (DRAM)
\item Network TCP layer, IP layer, Physical layer – LAN, Wi-Fi, WiMAX, 5G, 4G
\end{itemize}

Software:
\begin{itemize}
\item Application
\item OS Network Stack – Application layer
\end{itemize}

These use deep and wide multilevel architectures for compute and memory to accelerate performance.  A cache is commonly used as a local store for memory. System memory is accessed on an External I/O interface. We are seeing a wall being met in single thread and single process performance. We have consequently moved to multi-threaded and multi-process designs. However, these are hitting a wall due to the overhead of maintaining coherence and synchronization in a multilevel cache and memory. The features are implemented in the constraints of the enterprise vendor or customer environment.

This is further exacerbated through latencies in accessing external interfaces whether in an external multilevel cache, memory or I/O (network). This highlights a wall in enhancing single threaded / process implementations using deep compute architectures. Additionally, multi-threaded / process implementations are hitting a fundamental design barrier and wall using a local cache memory store which has to be coherent and synchronized across internal and external I/O.
  
\begin{figure}[!h]
  \caption{Conventional platform, multilevel architectures using wide, deep internal and external components.}
  \centering
  \includegraphics[width=0.8\columnwidth]{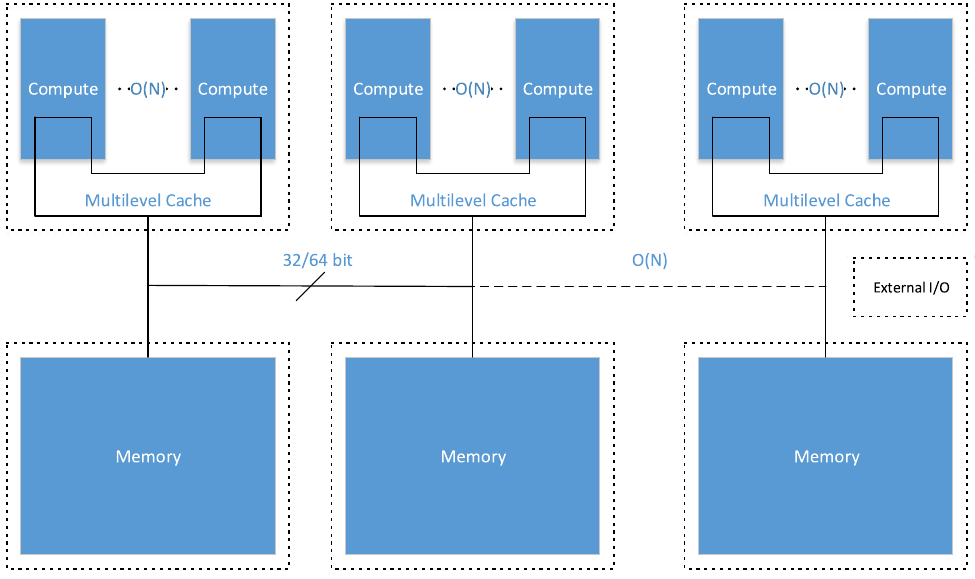}
  \label{fig:mlc}
\end{figure}

I/O access latency penalty (units)
\begin{itemize}
\item Internal 1 - 10’s 
\item External 100 - 1000’s 
\end{itemize}

Figure~\ref{fig:mlc} illustrates an architecture using a set of plug and play interfaces to scale performance. These are integrated in various topologies using a combination of internal and external I/O. Some common topologies include Ring, Mesh, Star and Ad hoc. Trade-off considerations around performance, price and technology constraints partition the design across internal and external components. The topologies are constrained in the limitations of the interfaces. Increasingly architectures are consolidating these hierarchies in a single technology using compute and multilevel caches with wide and deep configurations. 

\begin{figure}[!h]
  \caption{Producer Consumer.}
  \centering
  \includegraphics[width=0.6\columnwidth]{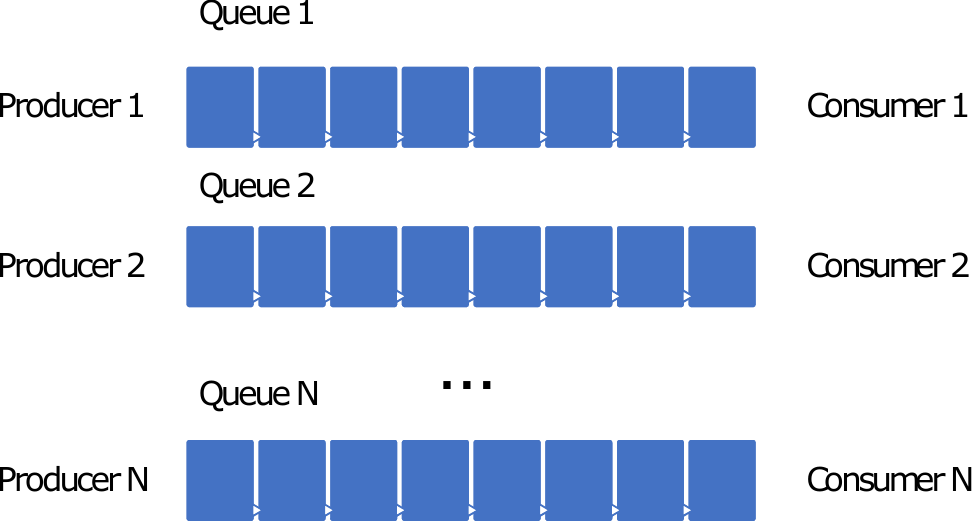}
  \label{fig:prodcons}
\end{figure} 

With the exponential growth (Moore’s Law) in technology development we find that developers have increased access to commoditized compute (RISC, CISC) technology. Platforms based on commodity Linux solution are widely deployed in the enterprise. Application developers are concerned about application performance and scalability in the cloud. Application performance bottlenecks are constantly evolving in a tiered internet. They vary around system constraints limitations in the kernel functionality. However, application scalability is bounded in fundamental constraints of application development arising from a producer consumer model. The Producer Consumer or Bounded buffer problem is an example of a multi-process synchronization challenge. It forms a central role in any Operating system design that allows concurrent process activity.

As we have N producers, N consumers and N queues in the application Figure~\ref{fig:prodcons} we can see that there are opportunities for the synchronization through the use of semaphores, deadlock avoidance and starvation. If we imagine infinite resources, then the producer continues writing to the queue and the consumer has only to wait till there is data in the queue. The dining philosopher’s problem is another demonstration of the challenges in concurrency and synchronization.

\begin{figure}[!h]
  \caption{Proposed high-level architecture with compute (RISC, CISC) coupled to memory, no caches.}
  \centering
  \includegraphics[width=0.6\columnwidth]{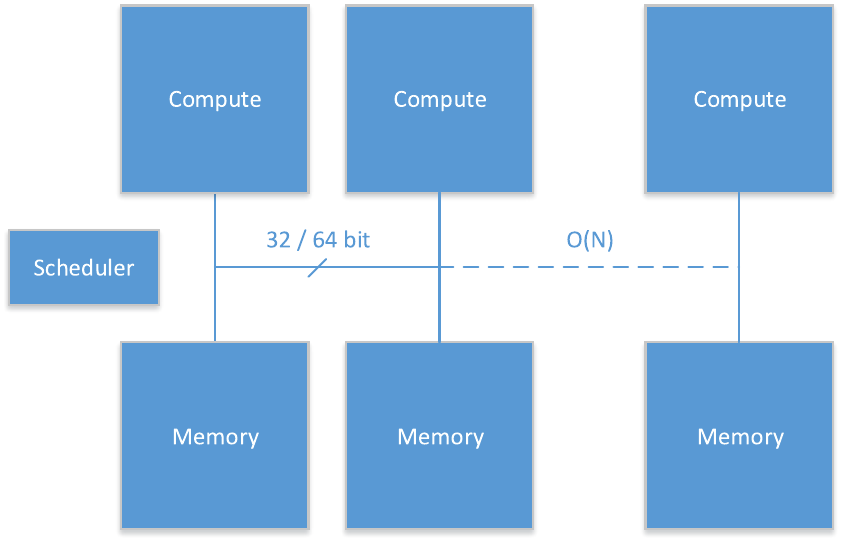}
  \label{fig:cmp}
\end{figure} 

\begin{figure}[!h]
  \caption{Proposed architecture with compute (RISC, CISC) coupled to memory using a crossbar switch, no caches. Architecture scales in underlying technology.}
  \centering
  \includegraphics[width=0.7\columnwidth]{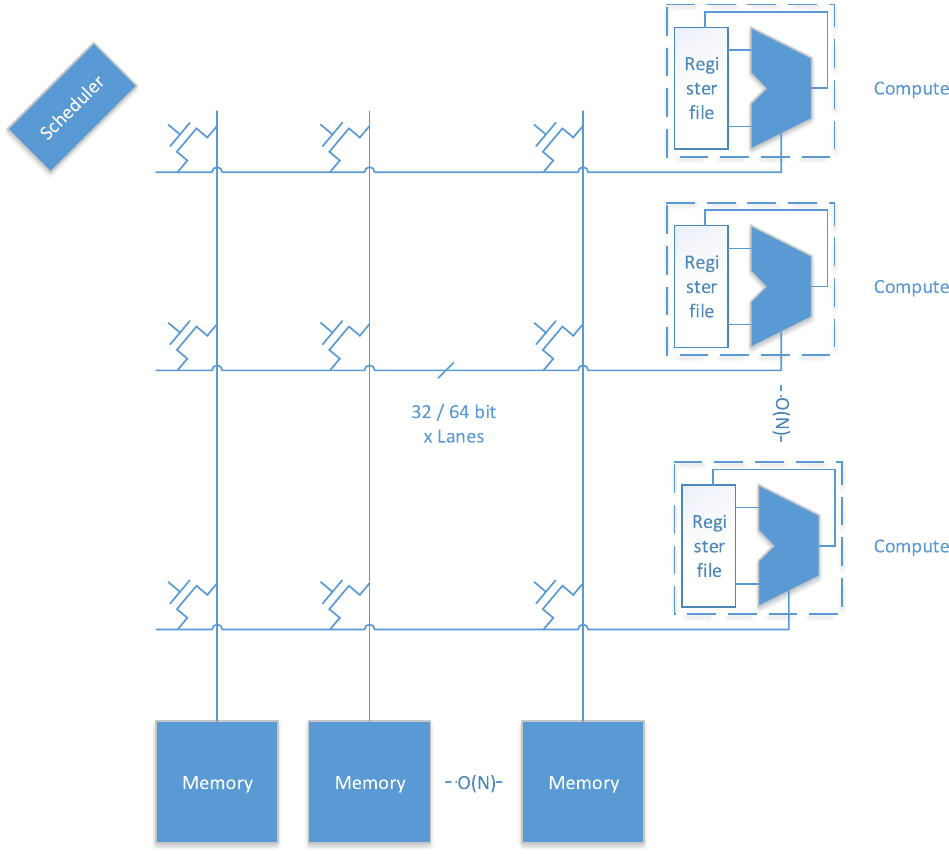}
  \label{fig:xbar}
\end{figure} 

Figure~\ref{fig:xbar} - The architecture addresses technology challenges in scaling the next generation internet including efficiency in the data center~\cite{Corbet:2005:LDD:1209083}. The architecture enables high performance commodity computing in the end user platform enabling technology of scale. Machine learning libraries like TensorFlow~\cite{abadi2016tensorflow} can be programmed on a homogeneous architecture fabric. This facilitates an abstraction for the development of algorithms~\cite{tensorflow2015-whitepaper} and software on an open platform using a host of programming languages~\cite{Hejlsberg:2010:CPL:1951915}.

Conventional computing platforms:
\begin{itemize}
\item Processor \textgreater 2GHz
\item Memory \textgreater 2GB
\end{itemize}

Proposed architecture (Figure~\ref{fig:xbar}):
\begin{itemize}
\item Network Bandwidth 1 – 10 Gbps
\end{itemize}

Shared bus implementation uses a Crossbar switch. These could use a buffer less design without any forwarding logic to access discrete memory banks using a bus select logic and multiple Lanes.  A scheduler is used to access individual memory banks on the system bus. Scheduler could use a static round robin scheduler or a priority request response implementation. 3D Stacking technology enables integration of heterogeneous technology integrating DRAM (\textgreater 1 GB) close to the Compute. 

System bottlenecks are constantly evolving. As infrastructure is increasingly being commoditized with a growth of development around open source technologies. It is essential that adequate bandwidth is provisioned in the cloud to allow for application scalability. Virtualization technology enables efficient partitioning of additional resources. As a metric, it is key to replicate scale the infrastructure maintaining redundancy to ensure quality of service in the end-to-end internet.

\section{Potential Benefits}

\begin{figure}[!h]
  \caption{Semantic web stack.}
  \centering
  \includegraphics[width=0.2\columnwidth]{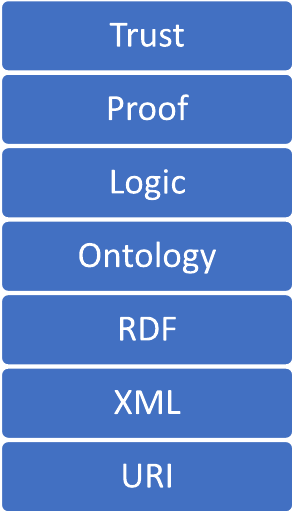}
  \label{fig:semantic}
\end{figure}

In the broader context of the internet it is always beneficial to host resources close to the client consumption including providing a larger bandwidth to the consumer. Additionally, open platforms and standards enable for a balanced distribution of available bandwidth resources allowing for a scalable platform for 10 billion consumers. Innovation and advancement are enabled through open source and open platforms around internet based wireless technology. The protocol stacks comprising the future semantic web data are as Figure~\ref{fig:semantic}.

\begin{itemize}
\item Web N. The internet is increasingly accessible to more than 10 billion users. It has been designed around Internet protocols and standards. The next generation of the web will use various Semantic web technologies.
\item Cloud computing. Rapidly commoditized infrastructure and Linux servers
\item Knowledge systems. Vast repositories of structured, unstructured data
\item Efficient programming languages. Github
\end{itemize}

Proposed architecture:
\begin{itemize}
\item Multiprocessor and SRAM memory tightly coupled – no caches
\item Generic configurable Network I/O – LAN, Wi-Fi, WiMAX, 5G, 4G
\item Programmable Network I/O - Network layers integrated in OS stack
\end{itemize}

The Stack machine  is a fundamental compute primitive. Processor and memory technology are now capable of integrating multi-Ghz and multi-Mb designs. There is a diminishing improvement for multi-Ghz processor designs as application memory is accessible in memory (DRAM).

\begin{figure}[!h]
  \caption{Proposed high-level architecture with stack machine tightly coupled to memory, no caches.}
  \centering
  \includegraphics[width=0.6\columnwidth]{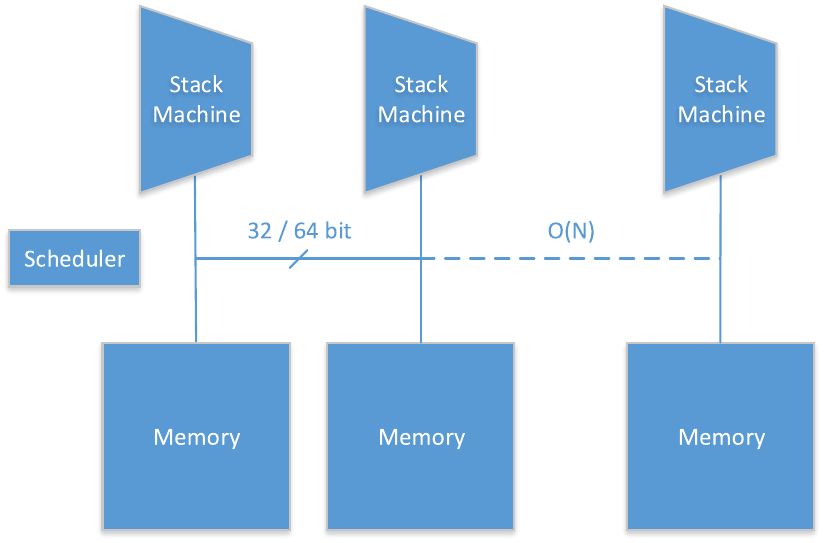}
  \label{fig:cmp_stack}
\end{figure} 

\begin{figure}[!h]
  \caption{Proposed architecture with stack machine tightly coupled to memory using a crossbar switch, no caches. Architecture scales in underlying technology.}
  \centering
  \includegraphics[width=0.7\columnwidth]{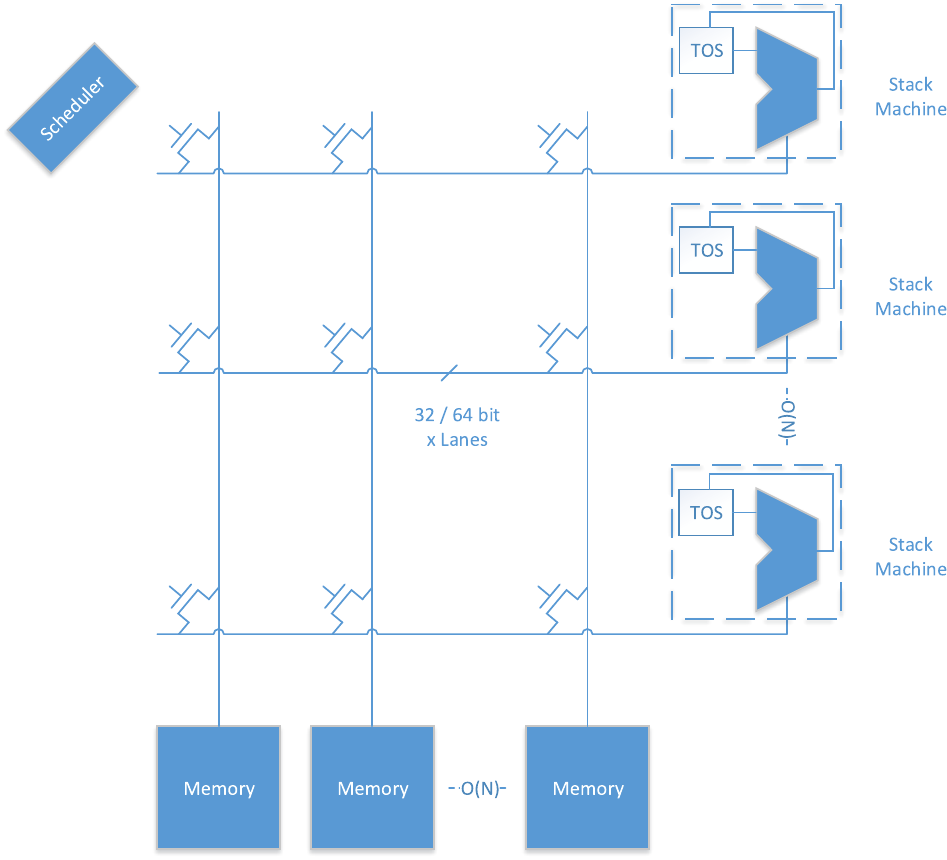}
  \label{fig:xbar_stack}
\end{figure} 
 
We propose a high-performance general-purpose web computing platform using a tightly coupled processor (no caches L1, L2) and memory (SRAM) - Figure~\ref{fig:xbar_stack}, Figure~\ref{fig:xbar}. A shared memory CMP architecture allows for a turn key, low cost solution to mobile connectivity allowing tight integration of processor technology and application specific software stack. A prototype device for mobile could contain 4 - 8 compute and 1 - 8 MB of SRAM memory~\cite{iandola2016squeezenet}. O(1) $\leftrightarrow$ O(N). Machine learning applications could be run efficiently on the platform~\cite{kumar2017resource}. 

As the device becomes accessible to more markets, we would see increasing accessibility to the internet~\cite{website:android}. Internet accessibility in a low-cost device enables scale in markets and applications. The platform uses general purpose processors in a shared memory environment to enable better programmability. Internet platforms enable open standards for technology development.

\begin{figure}[!h]
\centering
\begin{minipage}{.5\textwidth}
  \caption{g++ compiler.}
  \centering
  \includegraphics[width=0.5\columnwidth]{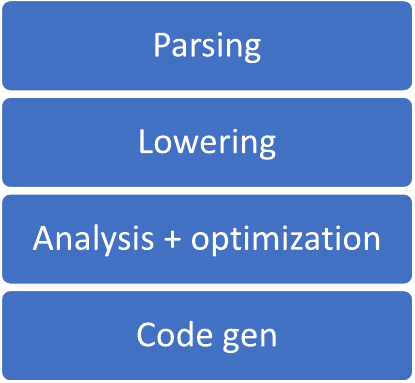}
  \label{fig:gpp}
\end{minipage}%
\begin{minipage}{.5\textwidth}
  \caption{V8 JS.}
  \centering
  \includegraphics[width=0.5\columnwidth]{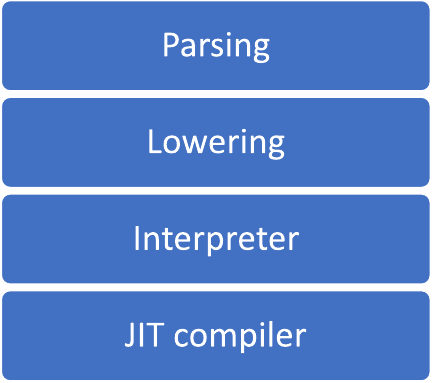}
  \label{fig:v8js}
\end{minipage}
\end{figure} 

Compilers and translators. Compiler is a Sequential batch architecture. Compilers translate information from one representation to another - Figure~\ref{fig:gpp}, Figure~\ref{fig:v8js}. Most commonly, the information is a program. Compilers translate from high-level source code to low-level code. Translators transform representations at the same level of abstraction. 

\begin{itemize}
\item Windows - 50 million LOC
\item Google internet services - 2 billion LOC
\end{itemize}

Some designs use a virtual machine Eg. JVM that runs on the target architecture. Increasingly designs are converging around web ecosystems and the JavaScript Developer frameworks using a JIT compiler. The proposed solution supports both native implementation and those using a virtual machine. 

\section{Business impact}

Programming languages are supported in a specific Software vendor stack Eg. C++ / C\# (Microsoft), Java (Oracle), Python / JS (Google) to create a community developer ecosystem.  These were conventionally developed around vendor specific platforms such as the PC Desktop, Mac or Mobile etc. 
 
\begin{figure}[!h]
\centering
\begin{minipage}{.4\textwidth}
  \caption{Converged memory stack.}
  \centering
  \includegraphics[width=0.8\columnwidth]{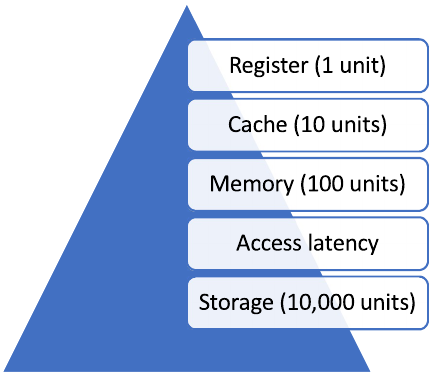}
  \label{fig:converged_mem}
\end{minipage}%
\begin{minipage}{.6\textwidth}
  \caption{Converged compute stack.}
  \centering
  \includegraphics[width=\columnwidth]{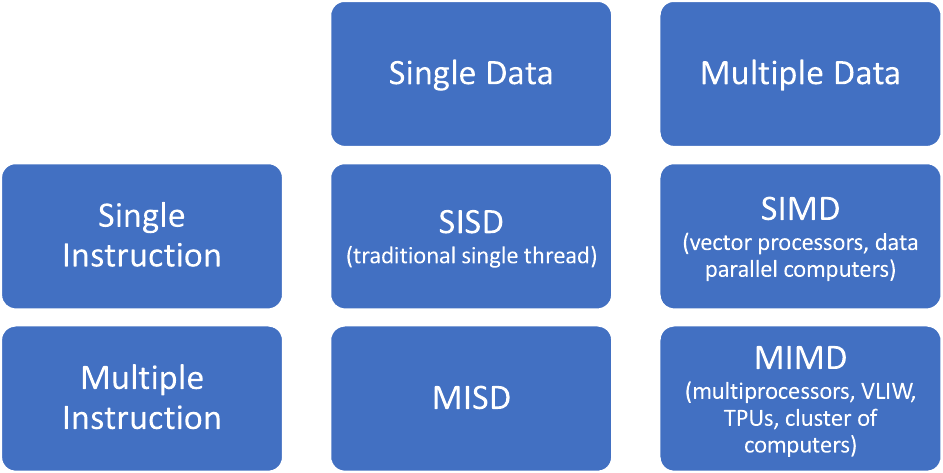}
  \label{fig:converged_compute}
\end{minipage}
\end{figure}

Figure~\ref{fig:converged_mem}, Figure~\ref{fig:converged_compute} show the converged memory and compute stacks in the Enterprise vendor software ecosystem. A specific combination of these primitive’s can be configured in the end-user application requirements such as networking or tensors~\cite{chollet2015keras}. These would use micro and macro consideration tradeoffs. Specific implementations of these designs are incentivized economically in the Software vendor ecosystems stack in a tiered technology industry leveraging a set of diversified business models.

\begin{figure}[!h]
  \caption{Proposed architecture stack.}
  \centering
  \includegraphics[width=0.7\columnwidth]{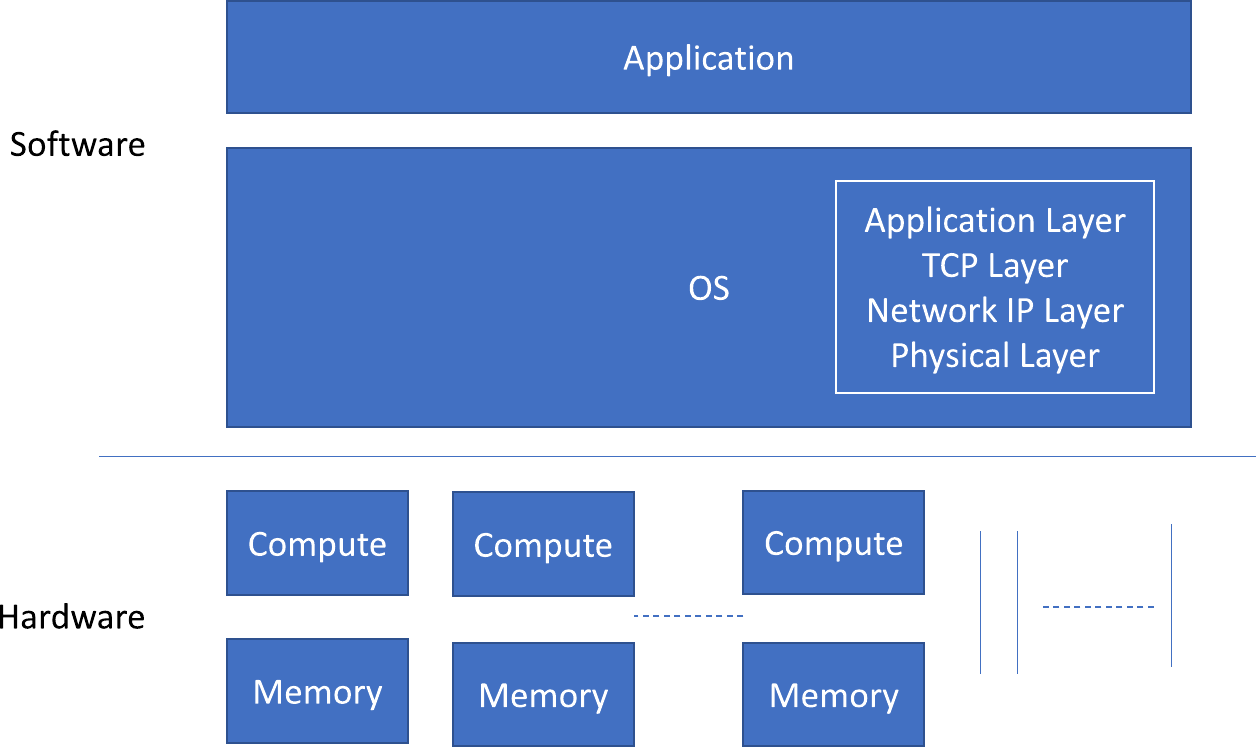}
  \label{fig:os_prop}
\end{figure}

Defined Platform architecture specification solution is agnostic to the underlying technology. The Programming language is the layer of abstraction. We facilitate an abstraction layer for the development of Software and Algorithms.

\begin{itemize}
\item Layered architecture
\item Pay as you go
\item Open Source
\item QOS requirement are guaranteed in the tiered Cloud Service provider
\item Interoperability
\item Scalability
\end{itemize}

We are seeing a broader industry wide convergence and disruption. The idea is a vital cog in the technology stack.

\begin{itemize}
\item Modus ponens, Conjecture
\end{itemize}

The architecture is a unified approach to bring next generation cognitive, low cost, mobile internet. The end user platform is able to scale as per the application requirements and network requirements in an efficient manner to improve cost, time to market, energy and accessibility - Figure~\ref{fig:os_prop}. It takes computing out of the data center and into end user platform enabling an internet of scale for the next century. Internet enables open standards, accessible computing and applications programmability on a commodity platform.

\begin{itemize}
\item Faster time to market
\item Increase quality and efficiency - Common Architecture Pattern 
\item Cost effective development of AI Solutions
\item On premise or Managed deployment
\item Lower cost of maintenance and support
\item Pluggable support for multiple Vendors
\end{itemize}

\section{Conclusion}

Increasing number of devices are being connected to the internet. The internet is an open platform for next generation technologies~\cite{Faulkner:17:H}. Open platforms enable better collaboration and innovation. The future of the internet is mobile as \textgreater 1 billion devices go online on IP. The presented architecture is a CMP design based on commodity processor and memory technology. We have an architecture with N (10's - 100's) compute connected to multi-MB SRAM memory using a shared memory system bus architecture. Number of compute and memory can scale in the power and performance requirements of the platform and the technology generation.

\section{Highlights}

\begin{itemize}
\item We propose an end to end architecture for the next generation web infrastructure.
\item Multiprocessor and SRAM memory tightly coupled - no caches. Shared memory system bus (crossbar switch). 
\item Homogeneous architecture. The Stack machine is a fundamental compute primitive. 
\item Applications can be developed and deployed on a multitude of host platforms.
\item A prototype device could contain N compute and N MB of memory. O(1) $\leftrightarrow$ O(N).
\item Broader industry wide convergence and disruption. Idea is a vital cog in the technology stack. 
\end{itemize}

\bibliographystyle{abbrv}
\bibliography{references} 

\end{document}